\documentclass[amsmath,amssymb,nofootinbib]{revtex4}

\pdfoutput=1

\usepackage{graphicx}
\usepackage{dcolumn}
\usepackage{bm}
\usepackage{epsfig}
\usepackage{color}
\usepackage{wasysym} 
\usepackage{hyperref}

\usepackage{aas_macros}

\hypersetup{
    colorlinks=true,
    linkcolor=red,
    citecolor=blue,
}

\def\fun#1#2{\lower3.6pt\vbox{\baselineskip0pt\lineskip.9pt
  \ialign{$\mathsurround=0pt#1\hfil##\hfil$\crcr#2\crcr\sim\crcr}}}
\def\simgt{\mathrel{\lower0.6ex\hbox{$\buildrel {\textstyle >}
 \over {\scriptstyle \sim}$}}}
\def\simlt{\mathrel{\lower0.6ex\hbox{$\buildrel {\textstyle <}
 \over {\scriptstyle \sim}$}}}

\def\bea{\begin{eqnarray}}
\def\eea{\end{eqnarray}}
\def\be{\begin{equation}}
\def\ee{\end{equation}}

\input epsf

\newcommand{\mpcoh}{\,h^{-1}\,{\rm Mpc}}

\def\be{\begin{equation}}
\def\ee{\end{equation}}
\def\ba{\begin{eqnarray}}
\def\ea{\end{eqnarray}}

\begin{document}

\preprint{}

\title{Cosmic Web and Environmental Dependence of Screening: Vainshtein vs. Chameleon}

\author{Bridget Falck$^{(1)}$, Kazuya Koyama$^{(1)}$, Gong-Bo Zhao$^{(2,1)}$}

\bigskip

\vskip1cm

\affiliation{
\vskip0.3cm
$^{(1)}$Institute of Cosmology \& Gravitation, University of Portsmouth, Burnaby Road, Portsmouth, United Kingdom \\
$^{(2)}$
National Astronomy Observatories, Chinese Academy of Science, Datun Road, Beijing, P. R. China
}


\begin{abstract}
Theories which modify general relativity to explain the accelerated expansion of the Universe often use screening mechanisms to satisfy constraints on Solar System scales. We investigate the effects of the cosmic web and the local environmental density of dark matter halos on the screening properties of the Vainshtein and chameleon screening mechanisms. We compare the cosmic web morphology of dark matter particles, mass functions of dark matter halos, mass and radial dependence of screening, velocity dispersions and peculiar velocities, and environmental dependence of screening mechanisms in $f(R)$ and nDGP models. Using the ORIGAMI cosmic web identification routine we find that the Vainshtein mechanism depends on the cosmic web morphology of dark matter particles, since these are defined according to the dimensionality of their collapse, while the chameleon mechanism shows no morphology dependence. The chameleon screening of halos and their velocity dispersions depend on halo mass, and small halos and subhalos can be environmentally screened in the chameleon mechanism. On the other hand, the screening of halos in the Vainshtein mechanism does not depend on mass nor environment, and their velocity dispersions are suppressed. The peculiar velocities of halos in the Vainshtein mechanism are enhanced because screened objects can still feel the fifth force generated by external fields, while peculiar velocities of chameleon halos are suppressed when the halo centers are screened.
\end{abstract}

\maketitle

\section{Introduction}

One of the most fundamental problems in modern cosmology is providing a theoretical explanation for the late-time acceleration of the Universe. The acceleration could be due to the presence of the mysterious ``dark energy'' or a sign that modifications to General Relativity (GR) are needed on cosmological scales. Modifications to GR generally introduce new degrees of freedom in the gravitational sector, which often generate an additional ``fifth force''. In the Solar System, this fifth force needs to be suppressed to pass stringent tests of gravity. Screening mechanisms have been developed recently to suppress the fifth force on small scales, which allows significant modifications to gravity on cosmological scales while satisfying local tests of gravity. For example, the chameleon mechanism makes the mass of the scalar field large in high density environments~\cite{Khoury:2003rn}, and the Vainshtein mechanism utilises non-linear derivative self-interactions to suppress the coupling to matter \cite{Vainshtein} (see Ref.~\cite{Joyce2014} for a review). 

The chameleon mechanism appears in $f(R)$ gravity, where the Einstein-Hilbert action is replaced by an arbitrary function of the Ricci curvature~\cite{fr}. A popular choice for this function is the model of Ref.~\cite{husawicki}, which can reproduce the background expansion of LCDM while satisfying Solar System tests of gravity. The chameleon mechanism becomes active for objects with a deep Newtonian potential, thus tests of gravity in this framework involve searching for objects with shallow potentials (and usually low densities) which are unscreened. 
The Vainshtein mechanism was discovered in the context of massive gravity but also appears in galileon cosmology and braneworld models~\cite{Koyama2011,Sbisa2012,Chow2009,Silva2009,DGP,massive,brane}. Since the Vainshtein mechanism depends on derivative self-interactions, it does not rely on a particular form of the scalar potential or couplings of the scalar to matter, as in the chameleon mechanism. However, Vainshtein screening does depend on the dimensionality of the source~\cite{Brax:2011sv,Bloomfield2014}, and therefore it operates differently for the halos, filaments, walls, and voids of the cosmic web~\cite{Falck2014}. There has also recently been an indication that the shape of the source may have an effect on the strength of chameleon screening~\cite{Burrage2014}.

The purpose of this paper is to systematically compare the chameleon and Vainshtein screening mechanisms, with particular focus on their cosmic web and environmental dependence. 
Due to the non-linear nature of screening mechanisms, $N$-body simulations have been developed to consistently solve the non-linear field equations in a cosmological background for specific modified gravity models, including $f(R)$ and the braneworld model of DGP~\cite{Oyaizu2008,Schmidt2009code,Zhao2011MLAPM,Li:2011vk,Li:2013nua,Li:2013tda}. 
Most studies have focused on a particular type of screening. Though comparisons have been made (see, e.g., Ref.~\cite{Schmidt:2010jr}), we present here tuned simulations to enable a direct comparison. 
We run $N$-body simulations that use the same initial conditions to minimise the cosmic variance, and with parameters tuned to have the same $\sigma_8$ at $z=0$ in both models to remove the effect of the linear growth of structure on the screening. 

In a previous paper~\cite{Falck2014}, we have showed that the Vainshtein mechanism operates differently according to the morphology of the cosmic web. We extend our analysis to the chameleon mechanism to investigate whether we can use this property to distinguish between the two mechanisms. In the chameleon mechanism, it was shown that the screening of dark matter halos depends on their mass while the suppression of the fifth force in the Vainshtein mechanism does not depend on the mass~\cite{Schmidt:2010jr}. We confirm this finding and study the radial dependence of the screening in both mechanisms. 
Another interesting feature of the Vainshtein mechanism is how, in contrast with chameleon, a screened body can still feel the fifth force generated by external fields as long as its wavelength is long compared to the Vainshtein radius~\cite{Hui2009,Hui:2012jb}. We confirm this picture by studying velocity dispersions and peculiar velocities of dark matter halos. Finally, we study the environmental dependence of the screening. In chameleon models, subhalos and small halos in dense environments tend to be screened by nearby halos~\cite{Zhao:2011cu,Cabre2012}. This environmental screening was a key property in developing astrophysical tests of the chameleon mechanism~\cite{Cabre2012,Vikram2013,Gronke2014,Terukina}. We will compare the environmental dependence of screening in the chameleon and Vainshtein mechanisms.

The paper is organised as follows. In section~\ref{sec:model}, we introduce two representative models for the chameleon and the Vainshtein mechanism. Then we describe our $N$-body simulations in detail. In section~\ref{sec:particles}, we study the fifth forces acting on dark matter particles. We briefly describe the ORIGAMI method of Ref.~\cite{Falck} to identify the morphology of dark matter particles and study the dependence of the screening mechanisms on the morphology of the cosmic web. In section~\ref{sec:halos}, we study dark matter halos. We first identify dark matter halos using the ORIGAMI code and study the screening of dark matter halos. We then study velocity dispersions and peculiar velocities of halos to investigate how screened bodies respond to external fields. Finally using the AHF code, we study the environmental dependence of screening on halos and subhalos. Section~\ref{sec:conclusion} is devoted to conclusions. 

\section{Models and Simulations}
\label{sec:model}

In this section, we introduce two representative models for the chameleon and Vainshtein mechanisms and describe our suite of $N$-body simulations. In the following and in the simulation codes, we use the quasi-static approximation, which ignores the time derivatives in the scalar field equations. Non-static simulations (without using this approximation) have been run for chameleon~\cite{Bose2015}, Vainshtein~\cite{WintherFerreira2015}, and Symmetron screening mechanisms~\cite{Llinares2013,Llinares2014}, and together with a detailed comparison of different codes~\cite{Winther2015Codes}, these studies have shown that the quasi-static approximation has a negligible effect on results.

\subsection{Model - Vainshtein mechanism}
In order to disentangle the effects of different cosmological backgrounds and those of the Vainshtein mechanism, we consider the normal branch Dvali-Gabadadze-Porrati (nDGP) braneworld model \cite{DGP} that has exactly the same expansion history as the LCDM model \cite{Schmidt:2010jr}. Under the quasi-static perturbations, the Poisson equation and the equation for the scalar field are given by \cite{Koyama:2007ih}
\begin{align}
\nabla^2 \Psi & = \nabla^2 \Psi_N + \frac{1}{2} \nabla^2 \varphi, \\
\nabla^2 \varphi & + \frac{r_c^2}{3 \beta(a) a^2}
[(\nabla^2 \varphi)^2 -(\nabla_i \nabla_j \varphi)(\nabla^i \nabla^j \varphi)]
= \frac{8 \pi G a^2}{3 \beta(a)} \rho \delta,
\label{eq:phievo1}
\end{align}
where $\delta$ is the density contrast, $\Psi$ is the Newtonian potential, and we define the Newtonian potential in GR as
\begin{equation}
\nabla^2 \Psi_N = 4 \pi G a^2 \rho \delta.
\end{equation}
The function $\beta(a)$ is given by
\begin{equation}
\beta(a) = 1 +  2 H r_c \left(1+ \frac{\dot{H}}{3 H^2} \right),
\label{eq:beta}
\end{equation}
where the cross-over scale $r_c$ is the parameter of the model. 
If we linearise the equations, the Poisson equation is given by
\begin{equation}
\nabla^2 \Psi = 4 \pi G a^2 \left(1+ \frac{1}{3 \beta(a)} \right) \rho \delta.
\end{equation}
Note that $\beta$ is always positive, so the growth of structure formation is enhanced in this model.

This model has one extra parameter, $r_c$, in addition to the usual cosmological parameters in the LCDM model. If $r_c$ becomes  larger, the enhancement of gravity is weaker and also the Vainshtein mechanism operates more efficiently as the amplitude of the non-linear terms are larger. Thus in this limit we recover LCDM.

\subsection{Model - chameleon mechanism}

As a representative model that exhibits the chameleon mechanism, we consider an $f(R)$ gravity model where the Einstein-Hilbert action is generalised to a function of the Ricci curvature. We consider a model where the function $f(R)$ is given by \cite{husawicki} 
\begin{equation}
f(R) = R - 2 \Lambda - f_{R0} \frac{\bar{R}^2}{R},
\end{equation}
where $\bar{R}$ is the present day background curvature. For the parameters we consider in this paper, $|f_{R0}| \leq 10^{-4}$, the background cosmology is well approximated as that in the LCDM model. 
Under the quasi-static approximations, the Poisson equation and the equation for the scalar field are given by 
\begin{align}
\nabla^2 \Psi &= \nabla^2 \Psi_N + \frac{1}{2} \nabla^2 \varphi,\\
\nabla^2 \varphi 
&=  \frac{a^2}{3} \delta R(\varphi)
+\frac{8 \pi G a^2}{3} \rho \delta.
\label{eq:phievo}
\end{align}
The fifth force $\varphi$ originates from the scalar field  $f_R \equiv  df/dR$ where $\varphi = f_R(R)- f_R(\bar{R})$. The potential $\delta R$ is defined as  $\delta R =R - \bar{R}$, which is a non-linear function of $\varphi$, realising the chameleon mechanism. If we linearise the equation, the scalar field equation becomes 
\begin{equation}
(\nabla^2 - a^2 \bar{\mu}^2 )\varphi = \frac{8 \pi G}{3} \rho \delta,
\end{equation}
where the mass $\bar{\mu}$ is determined by $\bar{R} $ and $f_{R0}$. Below the Compton wavelength $\bar{\mu}^{-1}$, the linearised equation for the gravitational potential becomes 
\begin{equation}
\nabla^2 \Psi = \frac{16  \pi G}{3} a^2 \rho \delta.
\end{equation}
On the other hand, above the Compton wavelength, we recover GR for the linearised perturbations. 

This model also has one extra parameter, $|f_{R0}|$, in addition to the usual cosmological parameters in the LCDM model. If $|f_{R0}|$ becomes smaller, 
the background Compton wavelength becomes shorter and also the chameleon mechanism becomes more efficient in recovering LCDM.

\subsection{Fifth force profiles in NFW halos}
\label{sec:NFW}

It is useful to construct analytic approximations for the scalar field solution inside dark matter halos. We assume the dark matter density profile is described by the NFW profile~\cite{Navarro:1996gj} with a mass $M_{\Delta}$. The mass is defined as the mass contained within the radius $r=r_{\Delta}$; the density at $r_{\Delta}$ is $\rho_{\rm crit} \Delta$, where $\rho_{\rm crit}$ is the critical density of the Universe. The NFW profile is given by
\begin{equation}
\rho(r) = \rho_s f \left(\frac{r}{r_s} \right), \quad
f(y) = \frac{1}{y (1+y)^2},
\end{equation}
where $\rho_s =\rho(r_s)$ is fixed so that the mass within $r_{\Delta}$ is $M_{\Delta}$. The scale radius $r_s$ is more conveniently parameterised by the concentration $c_{\Delta} = r_{\Delta}/r_s$. By integrating this density profile, we obtain the enclosed mass within the radius $r$, $M(<r)$, as 
\begin{equation}
M(<r) = M_{\Delta} \frac{F (c_{\Delta} r/r_{\Delta})}{F(c_{\Delta})}, \quad
F(y)= -\frac{y}{1+y} + \ln(1+y).
\label{NFWm}
\end{equation}

The scalar field equation in the nDGP model, Eq.~\ref{eq:phievo1}, can be solved analytically \cite{Schmidt:2010jr, Falck2014}   
\begin{equation}
  \frac{d\varphi}{dr} = \frac{G M(<r)}{r^2} \frac{4}{3 \beta} g\left(\frac{r}{r_*} \right), \quad
g(x) = x^3 \left( \sqrt{1+x^{-3}} -1 \right),
\end{equation}
where $r_*$ is the Vainshtein radius 
\begin{equation}
r_*= \left(\frac{16 G M(<r) r_c^2}{9 \beta^2}\right)^{1/3}.
\label{r*}
\end{equation}
Inside the Vainshtein radius, the scalar force is suppressed compared with the Newtonian potential due to the non-linear derivative interactions. Outside the Vainshtein radius, the linear solution is realised where $g(r/r_*) \to 1/2$. For a larger $r_c$, the Vainshtein radius is larger thus the region in which the fifth force is suppressed becomes larger and we recover GR. 

The scalar field in $f(R)$ gravity can be approximated as \cite{Schmidt:2010jr}
\begin{equation}
  \frac{d\varphi}{dr} =\frac{1}{3} \frac{G\Big( M(<r) - M(<r_{\rm scr}) \Big)}{r^2}
\end{equation}
for $r > r_{\rm scr}$ and $d\varphi/dr=0$ for $r < r_{\rm scr}$. The screening radius is obtained as \cite{Terukina}
\begin{equation}
r_{\rm scr} = \Big( \frac{2}{3} \frac{\Psi(r_{\Delta})} {|f_{R0}| F(c_{\Delta})} - \frac{1}{c_{\Delta}} \Big) 
r_{\Delta}, 
\end{equation}
where the Newtonian potential is given by $\Psi= G M_{\Delta}/r_{\Delta}$. Inside the screening radius, $r_{\rm scr}$, the scalar field is very massive and the fifth force is suppressed. Only the density outside $r_{\rm scr}$ contributes to the fifth force at $r > r_{\rm scr}$. The screening radius is determined by the ratio between the Newtonian potential of the halo and $|f_{R0}|$, thus it is mass dependent. See Ref.~\cite{frprofile} for detailed discussions about the scalar field profile inside halos. 

We define the ratio between the fifth force and the Newton force as
\begin{equation}
\Delta_M = \frac{F_5}{F_G}, 
\quad F_5 =\frac{1}{2} \frac{d \varphi}{ dr}, \quad F_G= \frac{d \Psi_N}{dr}.
\label{eqn:deltam}
\end{equation}
For linear solutions without screening, $\Delta_M=1/3 \beta$ in nDGP and $\Delta_M=1/3$ in $f(R)$.

\subsection{Simulations}

We perform $N$-body simulations for both nDGP and $f(R)$ models. To highlight the difference in the screening mechanism, we simulate pairs of nDGP and $f(R)$ models with identical $\sigma_8$ at $z=0$, which roughly removes the difference in screening due to the difference in linear growth. Specifically, we simulate three $f(R)$ models: F4 ($|f_{R0}|=10^{-4}$), F5 ($|f_{R0}|=10^{-5}$), and F6 ($|f_{R0}|=10^{-6}$), and three corresponding nDGP models whose parameters are listed in Table I. We also simulate the LCDM model to compare with. Though the F4 and F5 models may be in tension with constraints from the Solar System (see, e.g.,~\cite{husawicki}), the differences with respect to LCDM are more apparent in these models, thus they are useful for the purpose of probing the effect of the cosmic web morphology and local environment on the screening mechanisms.

\begin{table}[htdp]

\begin{center}
\begin{tabular}{c|c|c}

\hline\hline

$f(R)$                                   &       nDGP                          &                                       \\
\hline
F4: $|f_{R0}|=10^{-4}$                &   nDGP1: $H_0 r_c=0.57$              &    $\sigma_8=0.946$               \\
F5: $|f_{R0}|=10^{-5}$                &   nDGP2:  $H_0 r_c=1.20$             &   $\sigma_8=0.891$               \\
F6: $|f_{R0}|=10^{-6}$                &   nDGP3:  $H_0 r_c=5.65$             &   $\sigma_8=0.854$               \\

\hline\hline

\end{tabular}
\end{center}
\label{tab:param}
\caption{The parameters for $f(R)$ and nDGP simulations.}
\end{table}%

The baseline cosmology was chosen to be the model favoured by recent Planck observations \cite{Planck2013}: $\Omega_b h^2=0.022161,~\Omega_c h^2=0.11889,~\Omega_K =0,~h=0.6777,~n_s=0.9611$, and $\sigma_8=0.841.$ 
The simulations were performed using the AMR code of ECOSMOG \cite{Li:2011vk}, which has been used for simulations of $f(R)$ \cite{Li:2012by} as well as DGP and Galileon models~\cite{Barreira:2013eea, Li:2013tda, Li:2013nua}. The simulations use $256^3$ particles in a $L=64\mpcoh$ box from the initial redshift $z=49$ down to $z=0$. Each set of models ($f(R)$, nDGP, and LCDM) are simulated using the same initial condition, which is generated using MPGrafic~\cite{mpgrafic}, and we simulate three realisations for each model to reduce the sample variance. We use POWMES \cite{powmes} to measure the power spectrum.

\begin{figure}[ht]
  \centering{
  \includegraphics[width=\hsize]{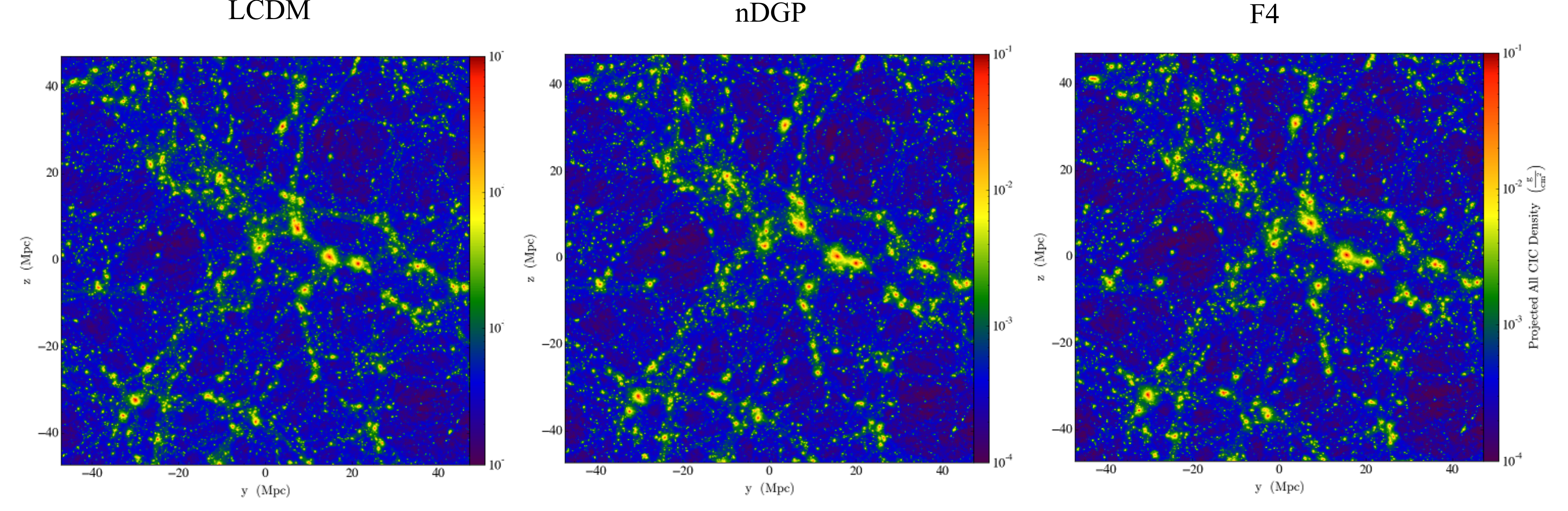}
  }
  \caption{The snapshots ($64\mpcoh\times64\mpcoh$) showing the projected density field at $z=0$ for LCDM {\it(left)}, nDGP1 {\it (middle)} and the F4 {\it (right)} models. 
   }
  \label{fig:snapshot}
\end{figure}

The simulations are visualised in Fig~\ref{fig:snapshot}, where the projected density field for LCDM and two MG models are shown at $z=0$. As shown, the structures are more clustered in nDGP and $f(R)$ models due to the enhanced gravity, although the enhancement of the clustering in these models are different.

\begin{figure}[ht]
  \centering{
  \includegraphics[width=17cm]{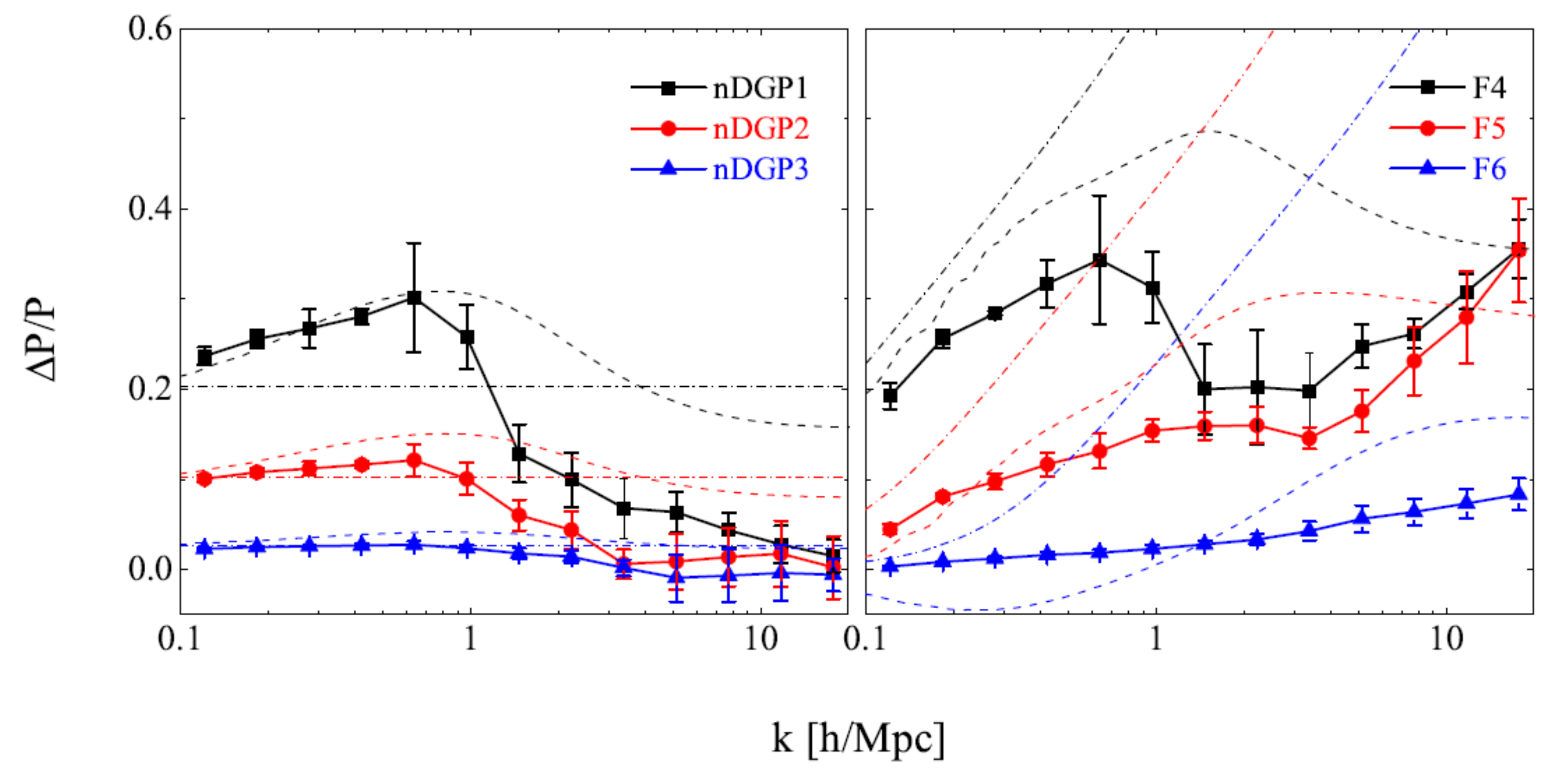}
  }
  \caption{The fractional difference in power spectrum for nDGP {\it (left)} and $f(R)$ {\it (right)} models with respect to the LCDM model. The data points with error bars show the simulation result, and the dashed (dash-dotted) curves show the Halofit (linear) predictions.  
   }
  \label{fig:power}
\end{figure}

To quantify the difference in clustering, we show the fractional difference of the power spectrum in $f(R)$ and nDGP models with respect to LCDM at $z=0$ in Fig.~\ref{fig:power}. The dotted lines show the linear prediction while the dash-dotted lines show the Halofit prediction~\cite{Smith:2002dz}.

An observation of Fig~\ref{fig:power} shows the following:

\begin{itemize}
\item Simulations of both models tend to agree with the linear prediction on large scales, and the agreement is better for the nDGP model. The excellent agreement with linear theory on large scales is one of the key features of the Vainshtein mechanism. This is because when the Vainshtein screening mechanism works, even if the fifth force is suppressed inside halos, these `screened' halos can still feel external scalar fields as long as those fields have wavelengths longer than the Vainshtein radius. We will confirm this picture later by studying the velocities of dark matter particles inside halos (see Section~\ref{sec:velocity}). However, in the $f(R)$ model with the chameleon screening mechanism, once the fifth force is suppressed inside dark matter halos, these screened halos no longer feel the external fifth force, making the screening happen on larger scales compared to the Vainshtein screening. This is why the linear perturbation theory fails on larger scales. 

\item Halofit overestimates the power for both models on nonlinear scales simply because Halofit, which is calibrated using LCDM simulations, does not have any screening mechanism built in. Note that though there are extensions of Halofit which include the chameleon screening, such as MGHalofit~\citep{MGHalofit}, we use Halofit to maintain consistency with the nDGP simulations. 

\item Halofit roughly captures the trend on quasi-nonlinear scales, and it works better for nDGP because it is less affected by the screening on these scales. 
\end{itemize}

\section{Dark Matter Particles}
\label{sec:particles}

In this section, we study the dependence of the chameleon and Vainshtein screening mechanisms on the cosmic web morphology of the dark matter particles. To measure the screening of individual particles, we calculate the ratio between the fifth force and gravitational force, $\Delta_M$ (Equation~\ref{eqn:deltam}), at the positions of the particles. If there are no non-linear interactions of the scalar field, then there is a linear relation between the fifth force and gravitational force: $\Delta_M=1/3$ in $f(R)$, and $\Delta_M=1/3 \beta$ in nDGP. However, if the screening mechanism is working, the fifth force will be suppressed relative to this linear relation. In Section~\ref{sec:screenpart}, we measure the deviation from this linear relation for each particle, $\Delta F$, such that a screened particle will have a vanishing fifth force and have $\Delta F\simeq -1$, while an unscreened particle will fall on the linear relation and have $\Delta F \simeq 0$. First we explain how we determine the cosmic web morphology of each particle and compare statistics from the various MG and LCDM simulations.

\subsection{ORIGAMI Morphology}

The cosmic web of large scale structure consists of an interconnected hierarchy of halos, filaments, walls, and voids. It is well described on large scales by the Zel'dovich approximation~\cite{Zeldovich:1969sb, Bond:1995yt} and forms naturally through the gravitational collapse of cold dark matter. Though there is a correlation between density and cosmic web morphology, the primary distinguishing feature of the different components of the cosmic web is the dimensionality of their collapse, which can be quantified e.g. by the relative values of the eigenvalues of the tidal tensor (see, e.g.,~\cite{Hahn:2006mk,Hoffman:2012ft,Cautun2013}). Here we use ORIGAMI\footnote{\url{http://icg.port.ac.uk/~falckb}} to determine the cosmic web morphology of each dark matter particle in a simulation~\cite{Falck}, unlike other methods based on the tidal tensor which define the cosmic web on a regular grid. ORIGAMI compares final positions to initial Lagrangian positions to determine whether a particle has undergone shell-crossing along a given set of axes (see Ref.~\cite{Falck} for details). Shell-crossing denotes the formation of caustics within which the velocity field is multi-valued, called the multi-stream regime. The number of orthogonal axes along which shell-crossing has occurred corresponds to the particle's cosmic web morphology and denoted by the morphology index $M$: halo particles have crossed along three axes ($M=3$), filaments along two ($M=2$), walls along one ($M=1$), and void particles are in the single-stream regime ($M=0$).

\begin{figure}[ht]
  \centering{
  \includegraphics[width=18cm]{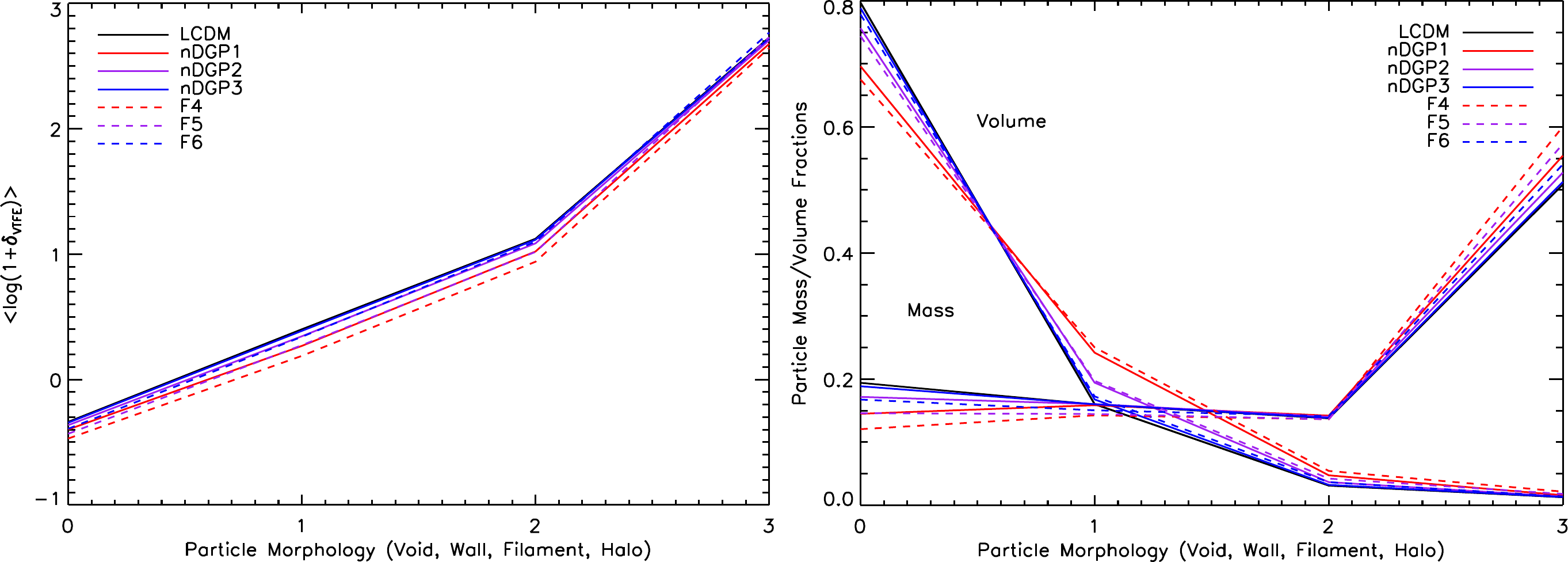}
  }
  \caption{{\it Left panel:} Average VTFE densities of void, wall, filament, and halo particles with morphology index $M=0$, 1, 2, and 3 respectively, for all 6 modified gravity models and LCDM. {\it Right panel:} Mass and volume fractions of particles according to their morphology.
  }
  \label{fig:origami}
\end{figure}

In Figure~\ref{fig:origami} we show the average densities and mass and volume fractions of the particles in all 7 simulations (3 $f(R)$, 3 nDGP, and LCDM) according to the ORIGAMI cosmic web morphology index $M$. (Note that all values for each model are calculated from the average over the three independent simulation runs.) The density for each particle is computed with the Voronoi Tessellation Field Estimate (VTFE)~\cite{Schaap:2000se, vandeWeygaert:2007ze}; the Voronoi tessellation partitions space into cells such that all points inside a particle's Voronoi cell are closer to that particle than to any other. The inverse volume of the Voronoi cell thus gives a scale-independent measure of the density at each particle location and is given by $\delta_{\rm VTFE} =\bar{V}/V-1$, where $V$ is the particle's Voronoi cell volume and $\bar{V}$ is the average of $V$ among all particles. The left panel of Figure~\ref{fig:origami} shows the average VTFE density of particles; as expected, the average density increases with morphology index such that voids have the lowest average densities and halos the highest. In general the particle densities of the modified gravity models are lower than in LCDM, for all particle morphologies, though there is hardly any change for halo particles.  When gravity is stronger, this can decrease average underdensities and increase overdensities, but the morphology of particles also changes; the void particles in LCDM that are walls or filaments in MG simulations are more likely to have higher density than other void particles, decreasing the average density of MG void particles, while the filament particles in LCDM that become halo particles in MG are likely to have lower density than other halo particles. The distributions of the densities change very little and are much broader than these small changes in the mean (see Figure 2 of~\cite{Falck2014}).

The right panel of Figure~\ref{fig:origami} shows the particle mass and volume fractions for each model as a function of morphology index. The mass fractions differ the most for void and halo particles, while the volume fractions differ most for void and wall particles (which have the largest volumes). The effect of the modified gravity model is to decrease the number (thus the mass) and volume of void particles with respect to LCDM, since gravity is stronger in these models and thus more shell-crossings occur. For the same reason, more of the volume is in the multi-stream regime with respect to LCDM, and more of the mass is in halos. For the mass and volume fractions and for the average densities, F4 and nDGP1 vary the most from the LCDM values, but the difference is greatest for F4.

\subsection{Screening}
\label{sec:screenpart}

The screening of each particle is given by the ratio of the fifth force, $F_5$, to gravitational force, $F_G$. This ratio has a linear theory value of $\Delta_M$ when screening is not effective: $\Delta_M=1/3$ in $f(R)$, and $\Delta_M=1/3 \beta$ in nDGP. We quantify the screening by calculating the deviation of the fifth force to gravitational force ratio from this linear relation,
\begin{equation}
\Delta F = \frac{F_5}{\Delta_M F_G} - 1,
\end{equation}
which ranges between $\Delta F = -1$ when screening is working to $\Delta F = 0$ when it is not. For some particles $\Delta F$ can be greater than 0 due to numerical noise, especially for low values of $F_5$ and $F_G$ (see, e.g.,~\cite{Moran2015}).

\begin{figure}[ht]
  \centering{
  \includegraphics[width=18cm]{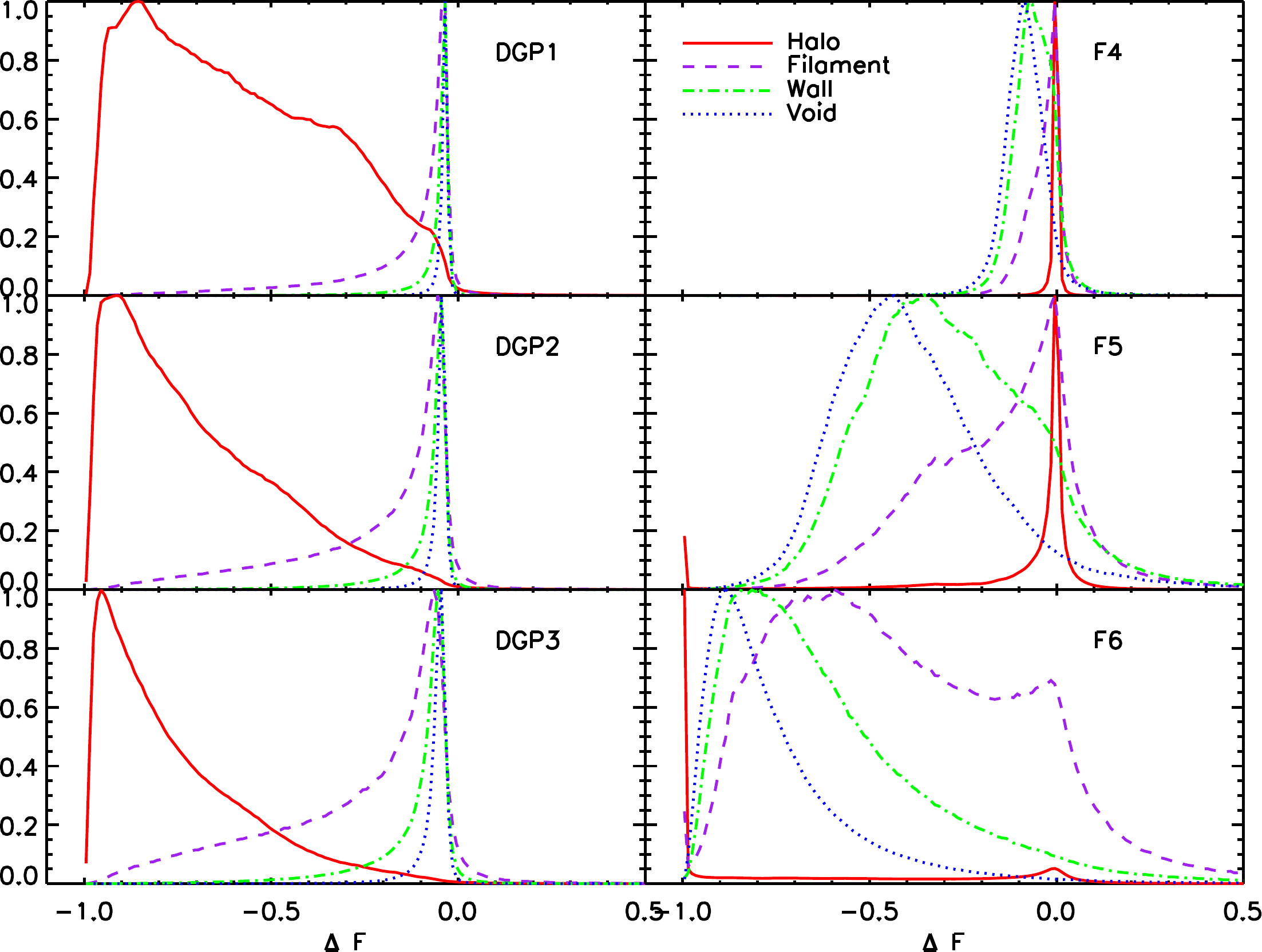}
  }
  \caption{Cosmic web morphology dependence of the Vainshtein {\it (left)} and chameleon {\it (right)} screening mechanisms, given by histograms of $\Delta F$, the deviation from the linear relation of the fifth force to gravitational force ratio. In DGP models with the Vainshtein mechanism, there is a clear difference between halo particles (solid, red line) and filament, wall, and void particles, while no such distinction exists for $f(R)$ models with the chameleon screening mechanism.
  }
  \label{fig:deltaF}
\end{figure}

In Figure~\ref{fig:deltaF} we show histograms of $\Delta F$ split according to ORIGAMI morphology for each modified gravity model. Note that the histograms are normalized to peak at unity so that the shape of each can be seen, so they do not reflect the relative abundances of the particles according to their cosmic web morphology; for that, refer to the mass fractions in Figure~\ref{fig:origami}. It is clear that, as we found in Ref.~\cite{Falck2014}, for all 3 nDGP models the halo particles are screened while the filament, void, and wall particles are unscreened. This reflects the dimensionality dependence of the Vainshtein mechanism~\cite{Brax:2011sv,Bloomfield2014}. The distribution of $\Delta F$ for halo particles is still broad, reflecting the fact that screening becomes weaker beyond the virial radius, so particles in the halos' outer edges can have larger values of $\Delta F$. 

On the other hand, there is no such morphology dependence of the screening mechanism for the chameleon models on the right side of Figure~\ref{fig:deltaF}. For the halos, screening is not very effective in F4, and most particles follow the linear relation; in F5 the halo particles have a small peak at $\Delta F = -1$ as the most massive halos become screened; and in F6 most of the halo particles are in screened halos, while some remain unscreened. Note that since the histograms are for halo particles and not halos themselves, massive halos are weighted more heavily, resulting in many screened halo particles in F6, while the smaller bump of unscreened halo particles is due to halos with low mass. We will look at the screening of halos in the next section. The wall and void histograms notably do not peak at $\Delta F = 0$ in F4, and the fifth force is further suppressed in F5 and especially F6. This is because the Compton wavelength is quite short, $\sim~1 {\rm Mpc}$ in F6, and the scalar field does not propagate beyond this length, providing a blanket screening for particles that are sparsely distributed (see Figure 2 of Ref.~\cite{Zhao2011MLAPM}). The filament distribution develops a double peak in F6: a narrow peak of unscreened filament particles, which have large forces and are in relatively dense environments, and a broader peak of low $\Delta F$ filament particles that are blanket screened.

\section{Dark Matter Halos}
\label{sec:halos}

We now turn to a comparison of the screening of halos in the Vainshtein and chameleon mechanisms. 
In Ref.~\cite{Falck2014} we found no dependence of the Vainshtein screening properties of halos on their morphological environment; since halo particles are effectively screened in the Vainshtein mechanism, it did not matter whether the halo was in a cluster or filament environment. For this study, we also found no dependence of the chameleon screening properties on the halo morphological environment, so in what follows we do not split halos into those in clusters, filaments, walls, and voids. 

After describing the halos and comparing mass functions, we will discuss the mass dependence of screening in Section~\ref{sec:mass}, the screening profiles in Section~\ref{sec:profiles}, the velocity dispersions and peculiar velocities in Section~\ref{sec:velocity}, and the environmental dependence of screening in Section~\ref{sec:environment}.

\subsection{Mass Functions}

We create a halo catalog by grouping together halo particles identified by ORIGAMI that are connected on the tessellation (see Ref.~\cite{Falck} for details). We use no density cut-off to define the halo edges, so ORIGAMI halos contain particles far outside the virial radius. To prevent over-connected halos, especially in dense environments, we first identify halo cores with particle densities above a VTFE density threshold $\rho/\bar{\rho}=150$. Note that any subhalos will be counted as part of the main halo, and we require that halos have a minimum of 20 particles. We define halo mass as $M_{200}$, the mass within $R_{200}$, beyond which the density drops below 200 times the critical density, $\rho_{\rm crit}$.

\begin{figure}[ht]
  \centering{
  \includegraphics[width=18cm]{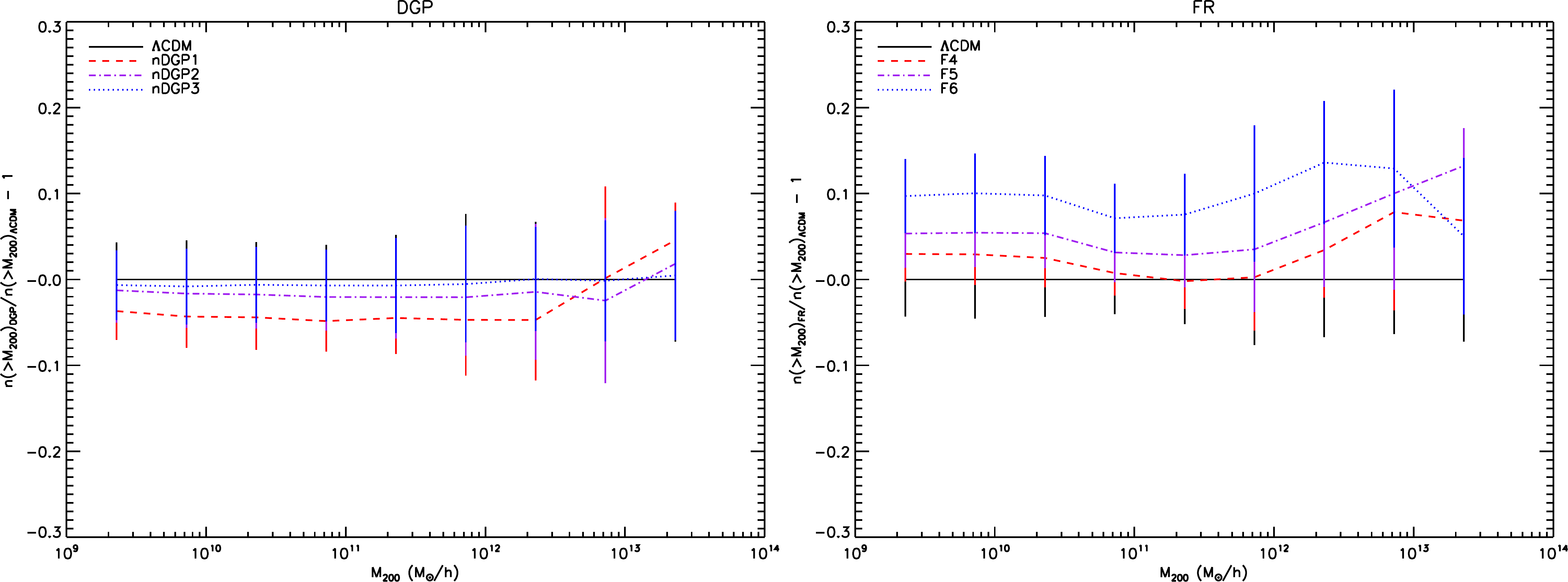}
  }
  \caption{Mass functions of nDGP {\it (left)} and $f(R)$ {\it (right)} models with respect to the LCDM mass functions. The lines are averages over the three independent simulation runs, and error bars represent the standard deviation.
  }
  \label{fig:massfn}
\end{figure}

Figure~\ref{fig:massfn} shows ratios of the cumulative mass functions of the nDGP and $f(R)$ models with respect to the LCDM mass functions. The mass functions for each of the three independent realizations are averaged, and error bars show the $1 \sigma$ variation. At high masses there are very few halos and the mass functions are dominated by cosmic variance noise, so masses above $4\times\,10^{13}\,h^{-1}\,{\rm M}_{\astrosun}$ are not shown.

The mass function ratios show that in both Vainshtein and chameleon screening, the model parameters with a stronger deviation from LCDM (nDGP1 and F4) result in a lower abundance of low mass halos than models with weaker deviation (nDGP3 and F6), but they have more halos at higher masses -- though with a small $64 \mpcoh$ box size, the statistics are poor for very high masses. This is because when gravity is stronger, such as in nDGP1 and F4, halos are more massive and small halos are absorbed into larger ones, leading to fewer low mass halos and more high mass halos. The difference between the DGP and $f(R)$ mass functions is that the DGP mass functions are suppressed with respect to LCDM, while the $f(R)$ mass functions are enhanced. As we will see in the next section, screening in the Vainshtein mechanism is effective at all masses, while screening does not operate for low mass halos in the chameleon mechanism, leading to a higher abundance of low mass halos in $f(R)$ compared to LCDM and nDGP. See Refs.~\cite{massfunction1, massfunction2, massfunction3, massfunction4} for analytic models of the mass function in these models.

\subsection{Screening - mass dependence}
\label{sec:mass}

\begin{figure}[ht]
  \centering{
  \includegraphics[width=18cm]{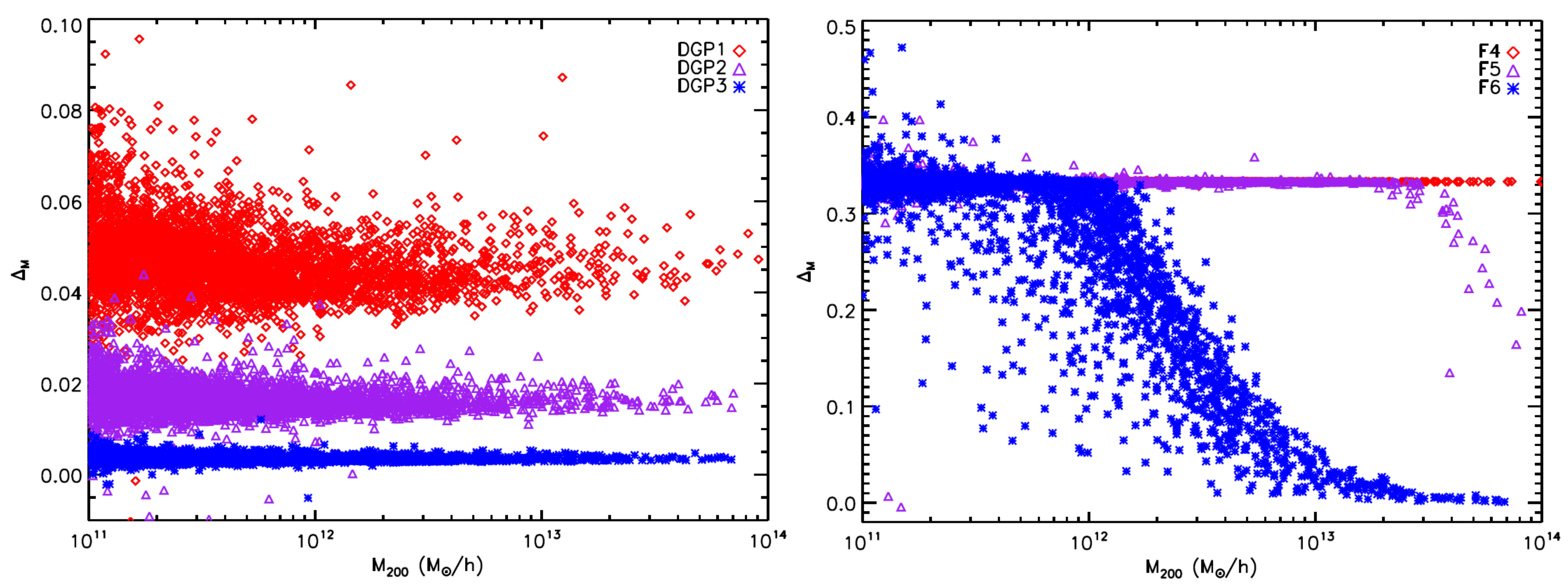}
  }
  \caption{Ratio of the fifth force to gravitational force as a function of halo mass for nDGP {\it (left)} and $f(R)$ {\it (right)} models.
  }
  \label{fig:dm_mass}
\end{figure}

As with the dark matter particles, to determine whether screening is effective we calculate the ratio of the fifth force to gravitational force, $\Delta_M$. The value of $\Delta_M$ for each halo is given by the average $\Delta_M$ of all the particles in the halo within the halo's virial radius, $R_{200}$. This is plotted as a function of halo mass, $M_{200}$, for all nDGP and $f(R)$ simulations in Figure~\ref{fig:dm_mass}.

For the DGP models, the $\Delta_M$ of the halos depends on the model parameter and is independent of mass; as the model parameter changes to make the deviation from LCDM stronger, $\Delta_M$ increases. However, note that all halos are screened in the Vainshtein mechanism: the linear values of $\Delta_M$ for nDGP1, nDGP2, and nDGP3 are 0.20, 0.11, and 0.03, respectively, well above the corresponding values in Figure~\ref{fig:dm_mass}. As we will see in the next section, the Vainshtein suppression gradually reduces (and $\Delta_M$ increases) outside the virial radius, but even including these particles in the calculation of $\Delta_M$ only increases $\Delta_M$ by $\sim 50 - 70$\%, and the halos remain screened overall. We note that although there is more scatter in $\Delta_M$ for nDGP1 and less for nDGP3, the scatter is about the same for each model in log-space (see Figure~\ref{fig:dgp_env}), so the larger scatter for nDGP1 is simply because the scatter is about a larger value.

In contrast to the Vainshtein mechanism, Figure~\ref{fig:dm_mass} shows there is a clear dependence on both mass and model parameter in the chameleon mechanism. When the deviation from LCDM is high, in F4, screening becomes ineffective for all halos; in F5, screening is effective only for high mass halos; and in F6, there is a population of unscreened small halos and a transition to screened large halos. Including particles outside the virial radius in the calculation of $\Delta_M$ has a very small effect, increasing $\Delta_M$ for some halos but not changing the overall trends; we will show in the next section that the radius of transition from screened to unscreened parts of the halo depends on the halo mass in the chameleon mechanism. The $\Delta_M$ scatter in F6 is due to the environmental dependence of screening, which we will discuss in Section~\ref{sec:environment}.

\subsection{Screening profiles}
\label{sec:profiles}

\begin{figure}[ht]
  \centering{
  \includegraphics[width=15cm]{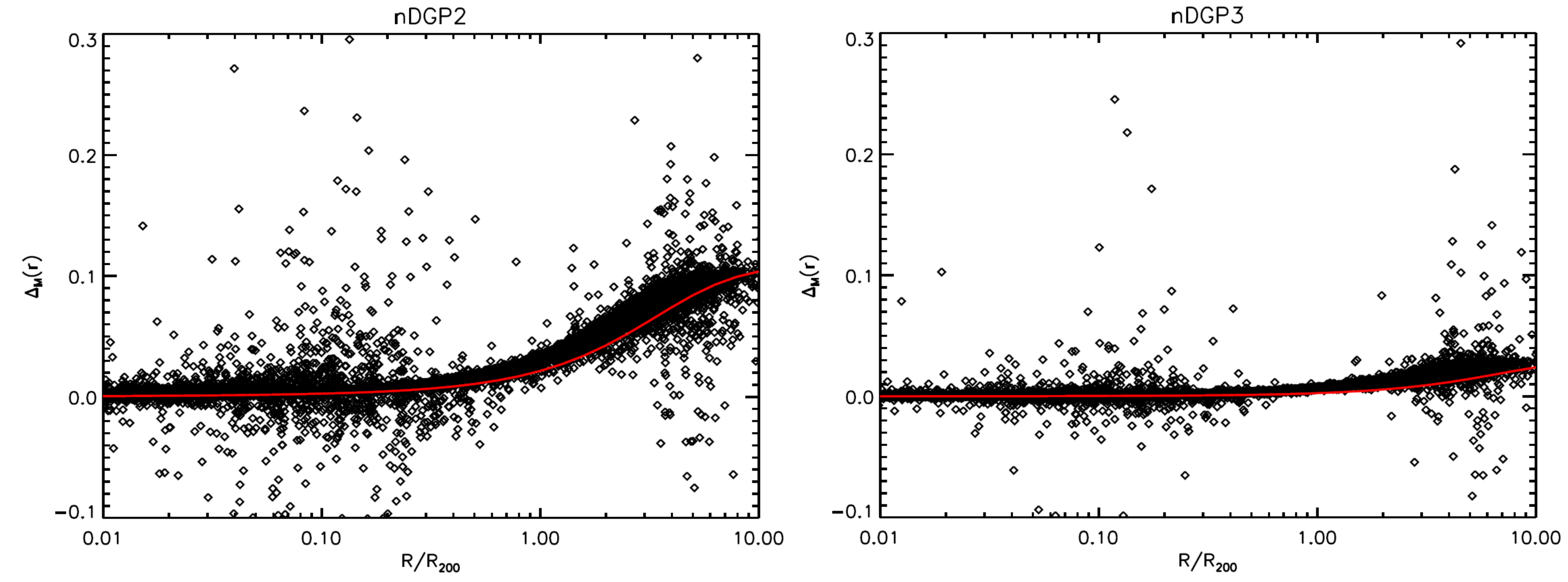}
  }
  \caption{Profiles of $\Delta_M$ in logarithmic bins of normalised radius, for nDGP2 {\it (left panel)} and nDGP3 {\it (right panel)}, with the analytic prediction for an NFW profile plotted in red. Screening profiles are independent of halo mass, and Vainshtein screening suppresses the fifth force within the virial radius.
  }
  \label{fig:dgp_profile}
\end{figure}

\begin{figure}[ht]
  \centering{
  \includegraphics[width=15cm]{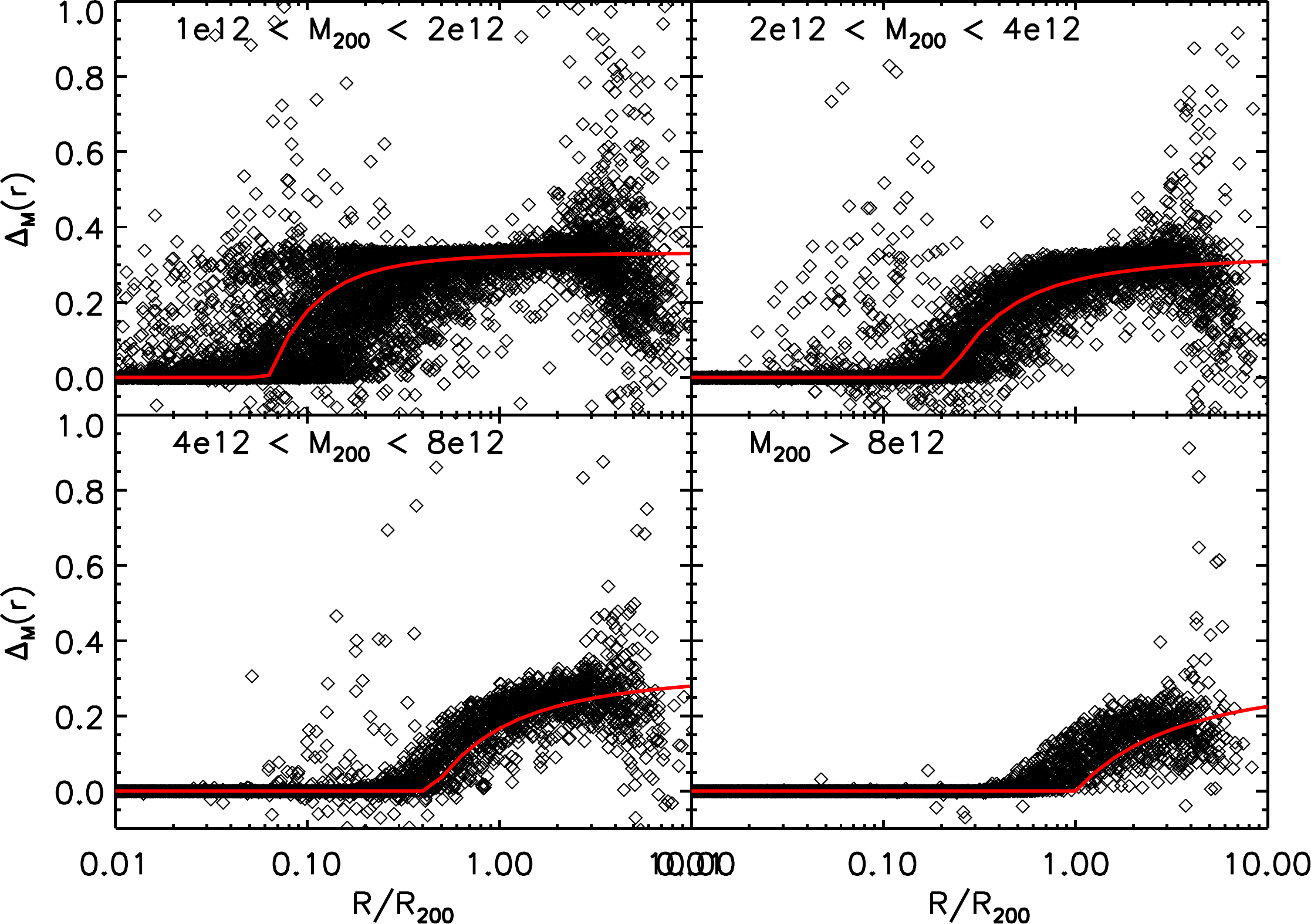}
  }
  \caption{Profiles of $\Delta_M$ in logarithmic bins of normalised radius for the F6 model, split into four bins of halo mass, with the analytic prediction calculated for each mass bin plotted in red. Screening profiles in the chameleon mechanism depend on halo mass.
  }
  \label{fig:fr_profile}
\end{figure}

To determine the radial dependence of the screening within the halos, here we calculate the ratio between the fifth force and Newtonian force as a function of normalised radius $R/R_{200}$ by averaging $\Delta_M$ of the halo particles in logarithmic radial bins. We only use halos with masses above $10^{12}\,h^{-1}\,{\rm M}_{\astrosun}$ to reduce numerical noise. $\Delta_M$ profiles are plotted in Figure~\ref{fig:dgp_profile} for nDGP2 and nDGP3. It is clear that regardless of halo mass, the Vainshtein screening profiles of dark matter halos are roughly the same and correspond quite well to the spherically symmetric analytic solution for an NFW profile (see Section~\ref{sec:NFW}), for which we assumed a concentration of 5 for all halos; we checked that the dependence on $c$ is quite weak. For both models, the fifth force is mostly suppressed within the virial radius while $\Delta_M$ increases outside the virial radius, and the magnitude of this increase is greater for models with a stronger enhancement to gravity, nDPG2 (and nDGP1, not shown).

In $f(R)$, Figure~\ref{fig:dm_mass} shows that chameleon screening does depend on mass, so in Figure~\ref{fig:fr_profile} we split up the $\Delta_M$ profiles into four different mass bins for the F6 model. In each mass bin, the analytic prediction (shown as the red line) again assumed $c=5$ for the NFW profile and was calculated for the average potential of the halos in the bin, except for the highest mass bin {\it (bottom right)} for which we excluded very high mass halos to calculate the average potential. There is a much sharper and less gradual transition from the chameleon screened inner regions of the halo to the unscreened outer regions compared to the Vainshtein mechanism, and the radius of this transition depends on the halo mass. There is again quite a good agreement with the analytic predictions from Section~\ref{sec:model}, given the large scatter in each mass bin and the simplifying assumptions made to calculate the predictions. Both Figure~\ref{fig:dgp_profile} and Figure~\ref{fig:fr_profile} highlight the importance of probing galactic halos in unscreened regions beyond their virial radii in order to detect deviations from LCDM, for all halo masses if the Vainshtein mechanism is operating and for high mass halos in the chameleon mechanism~\cite{Lam2012,Lam2013,Zu2014}.

\subsection{Velocities}
\label{sec:velocity}

In Ref.~\cite{Falck2014} we studied the super-imposability of field solutions in the Vainshtein mechanism, confirming the theoretical expectation that screened bodies can still feel the fifth force generated by external fields by looking at the velocity dispersions and peculiar velocities of dark matter halos. Here we repeat the analysis for the chameleon mechanism and directly compare the results for the two types of screening.

The non-linear derivative interaction that is responsible for the Vainshtein mechanism enjoys the Galilean symmetry~\cite{Nicolis:2008in}, which means that if the external fields have wavelengths that are long compared to the Vainshtein radius (Equation~\ref{r*}), we can regard the gradient of these external fields as constant gradients in the vicinity of an object. We can always add these constant gradients to the internal field generated by the object, and thus the internal field can superimpose with external fields. Even if the internal field is suppressed by the Vainshtein mechanism, the object still feels the fifth force generated by the external fields~\cite{Hui:2012jb}. 
On the other hand, in screening mechanisms that rely on non-linearity in the potential or the coupling function to matter, such as the chameleon and symmetron mechanisms, the internal field generated by an object does not superimpose with an external field. Therefore, the field inside the object loses knowledge of any exterior gradient and the fifth force generated by the external field, and thus once the object is screened, it does not feel any fifth force.

\begin{figure}[ht]
  \centering{
  \includegraphics[width=18cm]{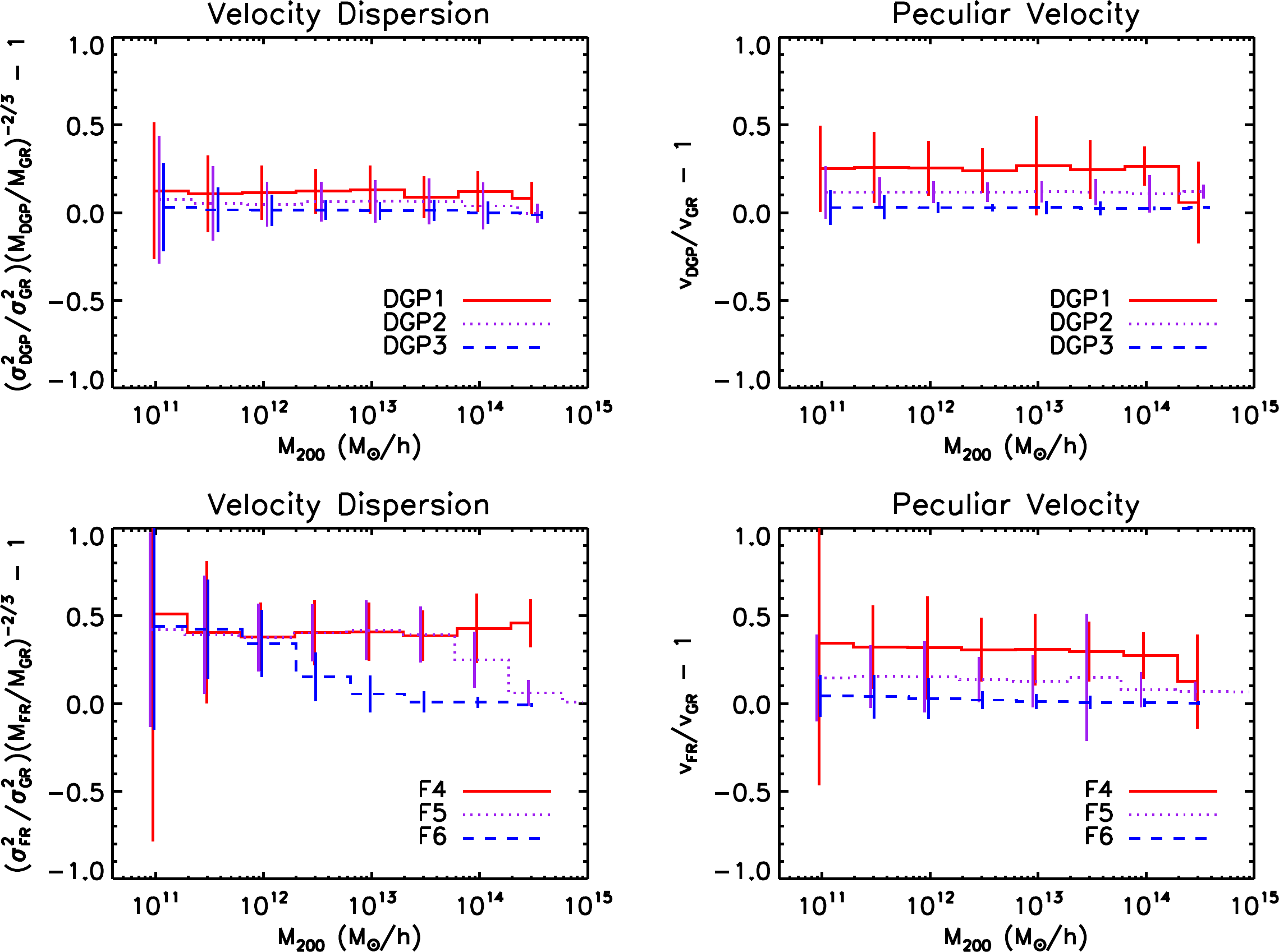}
  }
  \caption{Velocity dispersion {\it (left)} and peculiar velocity {\it (right)} ratios of nDGP {\it (top)} and $f(R)$ {\it (bottom)} models with respect to matched halos in the LCDM simulations, as a function of halo mass. Lines indicate the average value in each mass bin, and error bars indicate the scatter.
  }
  \label{fig:velocity}
\end{figure}

We study this by measuring the velocity dispersion and peculiar velocity of dark matter halos in the modified gravity simulations with respect to matched halos in the LCDM simulation. The enhancement of the velocity dispersion, measured using particles within $R_{200}$, with respect to LCDM is either suppressed or not depending on whether the screening is working, while the peculiar velocity, which is the magnitude of the average velocity of all halo particles, is enhanced with respect to LCDM if screening is not working or if the halo feels the linear fifth force induced by large scale structure. To match halos in the modified gravity simulations to those in the LCDM simulations, we compare halo centres and find the nearest MG halo to a given LCDM halo, where the halo centre is defined in ORIGAMI as the VTFE density-weighted average position of all halo particles. Both MG and LCDM halos must contain at least 100 particles, the MG halo mass must be at most 50\% greater than the LCDM halo mass, and the halos must be separated by less than 1$\mpcoh$.

In Figure~\ref{fig:velocity} we plot the velocity dispersion ratios and peculiar velocity ratios for all DGP and $f(R)$ models. The velocity dispersion ratios are scaled by the virial expectation, $\sigma^2\propto M^{2/3}$, to remove the standard mass dependence. The lines are the average values in bins of mass, and the error bars show the $1 \sigma$ standard deviation. Both the velocity dispersion and peculiar velocity ratios show no dependence on mass for the Vainshtein mechanism~\cite{Schmidt:2010jr,Falck2014}, but the peculiar velocity ratios deviate from 0, especially in the nDGP1 model, due to the effect of external fields. 

The $f(R)$ models, on the other hand, do show a mass dependence in the velocity dispersion ratios: in F4, where screening is not effective, the velocity dispersion is enhanced; in F5, the velocity dispersion is only suppressed at high masses; and in F6, the transition between enhanced and suppressed velocity dispersion occurs at an even lower mass. These trends mimic the mass dependence of the screening ratio, $\Delta_M$, in Figure~\ref{fig:dm_mass}. However, the peculiar velocity ratios do not show this mass dependence; they are suppressed in F6, somewhat enhanced in F5, and further enhanced in F4. Since most of the particles in a halo are located near the centre, the peculiar velocity effectively probes the halo centre and is suppressed because the halo core is screened, as seen in the $\Delta_M$ profiles for F6 in Figure~\ref{fig:fr_profile}. Unlike in the Vainshtein mechanism, once screened by the chameleon mechanism a halo does not feel the effect of external fields and so its peculiar velocity is suppressed. This casts doubt on the effectiveness of using redshift space distortions to detect $f(R)$ gravity at the level of F6. The Vainshtein mechanism is better suited to tests in the linear regime, since though halos are screened, they can still feel the effect of external fields induced by large scale structure.

\subsection{Screening - environment dependence}
\label{sec:environment}

In the previous sections, we have shown that chameleon screening depends on both the mass of the object and the model parameter, while Vainshtein screening is independent of halo mass (and see Ref.~\cite{Schmidt:2010jr}). In Section~\ref{sec:screenpart} and Ref.~\cite{Falck2014}, we showed that Vainshtein screening of dark matter particles depends on their cosmic web morphology; halo particles are screened while filament, void, and wall particles are not, reflecting the dimensionality dependence of the Vainshtein mechanism~\cite{Brax:2011sv,Bloomfield2014}. However, we found that the Vainshtein screening of halos themselves does not depend on their large scale cosmic web environment, and the chameleon mechanism has no cosmic web dependence for either particles or halos, so here we use a different definition of halo environment. 

We use a density-based definition of environment developed in Ref.~\cite{Haas2012},
\begin{equation}
D_{N,f} \equiv \frac{d_{N, M_N/M \geq f}}{r_N}.
\end{equation}
This is the distance $d_N$ to the $N$th nearest neighbor having mass, $M_N$, at least $f$ times as large as the halo mass, $M$, scaled by the virial radius of the neighboring halo, $r_N$. $D_{1,1}$ (hereafter just $D$) is almost uncorrelated with the halo mass~\cite{Haas2012}, and several studies have found that both chameleon and symmetron screening mechanisms correlate with this environment parameter~\cite{Zhao:2011cu,Cabre2012,Winther2012,Gronke2014,Moran2015}. In particular, while massive halos can be self-screened and in general smaller halos can be unscreened, small halos that live in dense environments (where $D$ is small) can be environmentally screened. Subhalos, especially those within the virial radius of their host halo with $D < 1$, are usually environmentally screened. Since ORIGAMI does not identify subhalos and found no halos with $D < 1$, here we use the AHF halo finder~\cite{Gill:2004km,Knollmann:2009pb} to study the environmental dependence of chameleon and Vainshtein screening. Note that the AHF and ORIGAMI virial mass functions for these simulations are similar, but a key difference between the two methods is that ORIGAMI halos end at their outer caustic or turn-around radius while AHF halos end at their virial radius, so ORIGAMI is useful for probing the unscreened outer regions of halos.

\begin{figure}[ht]
  \centering{
  \includegraphics[width=18cm]{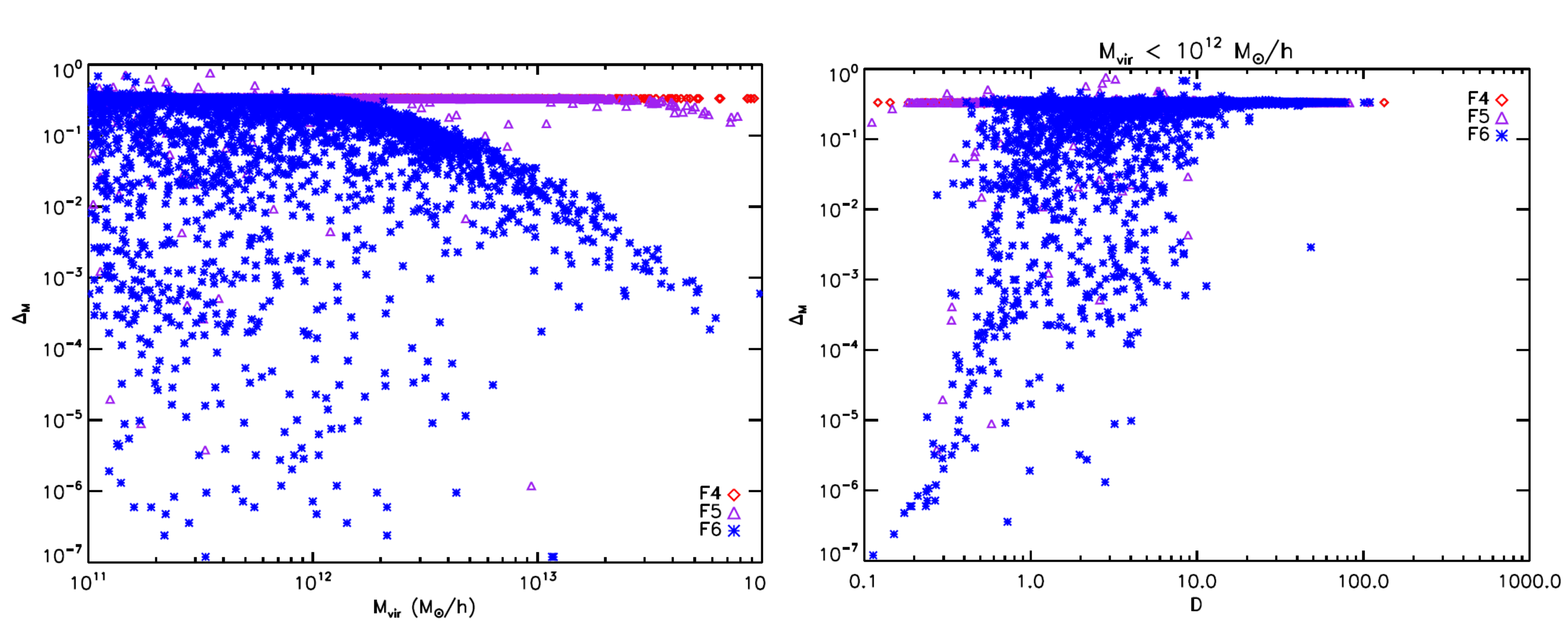}
  }
  \caption{The ratio of the fifth force to gravitational force for AHF halos in $f(R)$ models as a function of mass {\it (left panel)} and environment $D$ {\it (right panel)}. Only halos with mass below $10^{12}\,h^{-1}\,{\rm M}_{\astrosun}$ are shown in the right panel to remove halos that are massive enough to be self-screened. Chameleon screening displays both a mass and environment dependence.
  }
  \label{fig:fr_env}
\end{figure}

In Figure~\ref{fig:fr_env} we show the results for chameleon screening. In the left panel we plot $\Delta_M$ vs. halo mass, $M_{\rm{vir}}$, with a logarithmic scaling of $\Delta_M$ instead of the linear scaling of ORIGAMI values in Figure~\ref{fig:dm_mass}. The same trend in mass dependence of screening is seen, except AHF halos can have much lower values of $\Delta_M$, which we believe is due to the presence of subhalos. In the right panel of Figure~\ref{fig:fr_env} we show the environmental dependence of chameleon screening, after first removing halos with mass above $10^{12}\,h^{-1}\,{\rm M}_{\astrosun}$ that are self-screened. The trend is similar to that found by Ref.~\cite{Zhao:2011cu}: halos in dense environments, and especially those with $D<1$, tend to be screened while those in underdense environments, especially those with $D > 10$, are unscreened.

\begin{figure}[ht]
  \centering{
  \includegraphics[width=18cm]{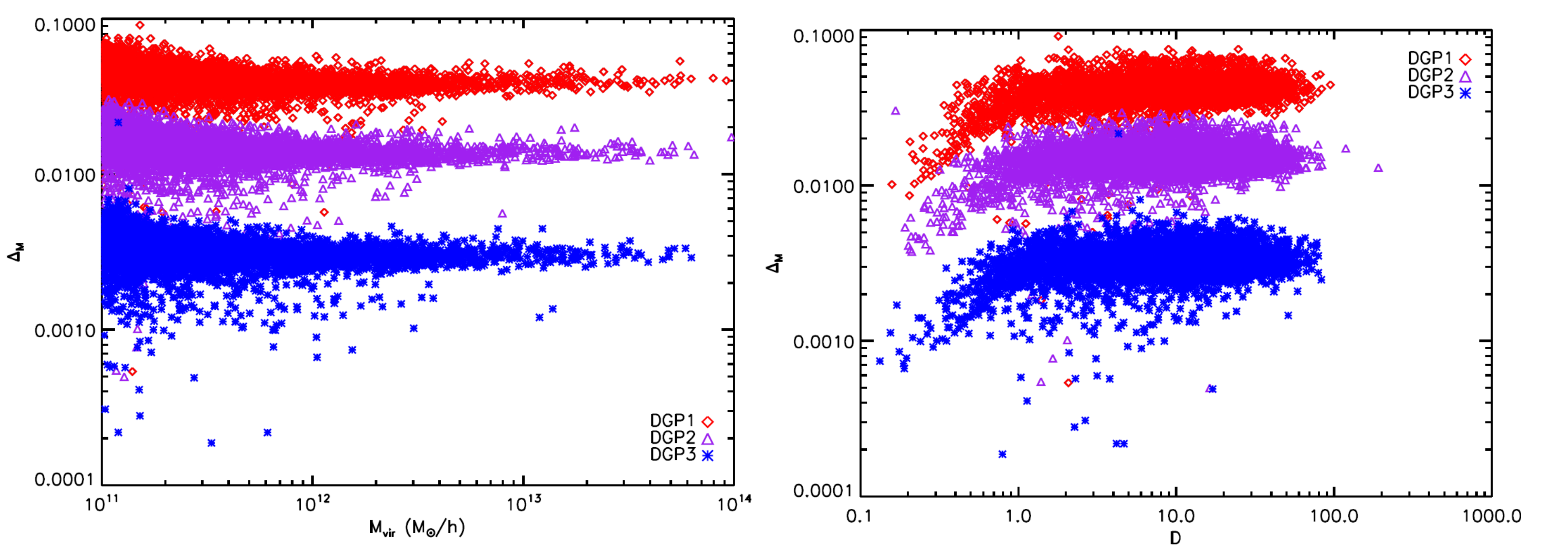}
  }
  \caption{The ratio of the fifth force to gravitational force for AHF halos in nDGP models as a function of mass {\it (left panel)} and environment $D$ {\it (right panel)}. There is no mass or environmental dependence of screening in the Vainshtein mechanism, and sub-halos within their host virial radius (with $D < 1$) have fifth forces that are even further suppressed.
  }
  \label{fig:dgp_env}
\end{figure}

The results for the Vainshtein screening mechanism are shown in Figure~\ref{fig:dgp_env}. The left panel shows the screening as a function of mass, with again no mass dependence, though there is a lot of scatter in $\Delta_M$ for low mass halos. The right panel shows $\Delta_M$ as a function of environment: in general there is no environmental dependence, but halos within their host halo with $D < 1$ are screened {\it even further} by their host halo. Note again that all halos are screened in the Vainshtein mechanism: the linear values of $\Delta_M$ for nDGP1, nDGP2, and nDGP3 are 0.20, 0.11, and 0.03, respectively, well above the corresponding values in Figure~\ref{fig:dgp_env}. The Vainshtein mechanism is therefore very efficient at screening halos: there is no dependence on mass~\cite{Schmidt:2010jr,Falck2014}, local cosmic web environment (i.e., halos in filaments vs. clusters)~\cite{Falck2014}, or density of their local environment.

\section{Conclusion}
\label{sec:conclusion}

We have presented a direct comparison of the effect of the Vainshtein and chameleon screening mechanisms on nonlinear structure formation using a suite of cosmological $N$-body simulations. We have simulated three pairs of $f(R)$ and nDGP models, with their parameters tuned so that each pair has the same $\sigma_8$ at $z=0$ to roughly remove any differences due to the linear growth of structure. We also simulate the LCDM model to have a baseline for comparison, and all simulations have the same initial conditions.

We found that while the Vainshtein mechanism depends on the cosmic web morphology of dark matter particles, first presented in Ref.~\cite{Falck2014}, there is no such dependence for the chameleon mechanism, potentially providing a new way of distinguishing between these two different types of screening. We confirmed that the screening of halos and halo profiles have no mass dependence in the Vainshtein mechanism, while both halos and halo profiles have a strong mass dependence in the chameleon mechanism.  To study the screening profiles of halos, it is especially important to have a halo-finder that identifies the outer regions of halos beyond the virial radius, where the fifth force of all halos in the Vainshtein mechanism and of massive halos in the chameleon mechanism is no longer suppressed. This means that observational tests of gravity are more likely to detect deviations from general relativity if they probe the outer regions of halos and clusters where screening is not effective.

This study presents the first test of the environmental dependence of screening in the Vainshtein mechanism. We found that Vainshtein screening of halos does not depend on their local environmental density, except for subhalos within the virial radius of their host halo, for which the screening is even stronger. Halos are thus very effectively screened in the Vainshtein mechanism, regardless of their mass, cosmic web environment, or local environmental density. In contrast, small halos in dense environments can be environmentally screened in the chameleon mechanism, while small halos in underdense environments are not screened, and large halos can be self-screened.

To study how screened bodies in the chameleon and Vainshtein mechanisms respond to external fields, we measured the velocity dispersions and peculiar velocities of MG halos with respect to matched halos in the LCDM simulations. In the Vainshtein mechanism, the velocity dispersion ratios are suppressed because screening works independent of halo mass, while the peculiar velocities are enhanced by the linear fifth force induced by large scale structure (and see~\cite{Falck2014}). In the chameleon mechanism, the velocity dispersion ratios match well the mass dependence of $\Delta_M$, where low mass halos in F6 are unscreened, but the peculiar velocity ratios in F6 are suppressed, since the centers of halos are screened and screened bodies do not feel the effect of external fields. This result suggests that it will be difficult to detect chameleon modified gravity using redshift space distortions -- even if a low mass halo is unscreened on average, its peculiar velocity will be suppressed if its centre is screened -- but that it is possible to detect models of gravity that employ the Vainshtein mechanism.

There has been interesting recent work studying the properties of voids in $f(R)$ gravity~\cite{Li:2011pj,Cai2014,Zivick2014}, and an obvious avenue of future work is to study voids in the Vainshtein mechanism, since halos are so effectively screened. Using ORIGAMI we have found that void particles, along with wall and filament particles, are unscreened in the Vainshtein mechanism; however, voids in ORIGAMI are defined as being in the single-stream regime (having undergone no shell-crossing), and it has been found that single-stream regions percolate, spanning the simulation volume~\cite{FalckVoids}. This means that walls and multi-stream regions do not completely surround voids, so ORIGAMI is not an optimal method to study voids found in the density field. It remains to be seen whether the lack of Vainshtein screening for dynamically defined void particles will hold for voids defined as underdensities, and this is the subject of future work.

\acknowledgments
BF and KK are supported by the UK Science and Technology Facilities Council (STFC) grants ST/K00090/1 and ST/L005573/1. GBZ is supported by the 1000 Young Talents program in China and by the Strategic Priority Research Program ``The Emergence of Cosmological Structures'' of the Chinese Academy of Sciences, Grant No. XDB09000000. Numerical computations were done on the Sciama High Performance Compute (HPC) cluster which is supported by the ICG, SEPNet, and the University of Portsmouth.

\bibliographystyle{JHEP}
\bibliography{refs}

\end{document}